\documentclass[12pt,a4paper]{article}
\pdfoutput=1

\let\counterwithin\relax
\usepackage[T1]{fontenc}
\usepackage[utf8]{inputenc}
\usepackage[width=0.95\textwidth, format=hang,justification=raggedright,labelfont=bf]{caption}
\usepackage{natbib}
\usepackage{enumitem}
\usepackage{booktabs}
\usepackage[all]{nowidow} 
\usepackage{amsmath,amssymb,amsfonts,amsthm}
\usepackage{tikz} 
\usepackage{tikz-qtree} 
\usepackage{xcolor}
\usepackage{cooltooltips}
\usepackage{MnSymbol}
\usepackage{graphicx}                           
\usepackage{microtype}

\usepackage{placeins} 
\usepackage{indentfirst} 
\usepackage{lscape}
\usepackage{rotating}
\usepackage[pdfborderstyle={/S/U/W 1},pdftex,breaklinks,colorlinks]{hyperref}
\AtBeginDocument{%
	\hypersetup{
		citecolor=blue,
		urlcolor =blue,
		menucolor = black,
		linkcolor = black}}
\usepackage{tocloft}

\usepackage{chngcntr} 
\counterwithin{figure}{section}
\counterwithin{table}{section}

\usepackage{algorithm}
\usepackage{algpseudocode}
\usepackage{graphics}
\usepackage{epsfig}
\usepackage{lscape}

\usepackage[left=2.5cm,right=2.5cm,top=2.5cm,bottom=2.5cm]{geometry}

\makeatletter
\newcommand*\bigdot{\mathpalette\bigdot@{.5}}
\newcommand*\bigdot@[2]{\mathbin{\vcenter{\hbox{\scalebox{#2}{$\m@th#1\bullet$}}}}}
\makeatother

\usepackage{listings}
\usepackage{xcolor}
\title{$K$-expectiles clustering  \thanks{Financial support of the European Union's Horizon 2020 research and innovation program "FIN-TECH: A Financial supervision and Technology compliance training programme" under the grant agreement No 825215 (Topic: ICT-35-2018, Type of action: CSA), the European Cooperation in Science \& Technology COST Action grant CA19130 - Fintech and Artificial Intelligence in Finance - Towards a transparent financial industry, the Deutsche Forschungsgemeinschaft's IRTG 1792 grant, the Yushan Scholar Program of Taiwan, the Czech Science Foundation's grant no. 19-28231X / CAS: XDA 23020303 are greatly acknowledged. All correspondence may be addressed to the authors by e-mail at \href{mailto:amyli999@hotmail.com}{amyli999@hotmail.com}.}\\ }

\date{February, 2021}

\author{Bingling Wang\thanks{ Humboldt-Universität zu Berlin, IRTG 1792, Dorotheenstr.1, 10117 Berlin, Germany, E-mail: bingling.wang@hu-berlin.de} \and Yinxing Li\footnotemark[2]  \thanks{Wang Yanan Institute for Studies in Economics, Xiamen University, 422 Siming S Rd,  361005 Fujian, China} \and Wolfgang Karl Härdle \footnotemark[2] \footnotemark[3] \thanks{ Humboldt-Universität zu Berlin, Blockchain Research Center, Unter den Linden 6, 10099 Berlin, Germany} \thanks{Sim Kee Boon Institute for Financial Economics,  Singapore Management University, 50 Stamford Road, Singapore 178899} \thanks{ Faculty of Mathematics and Physics, Charles University, Ke Karlovu 3, 121 16 Prague, Czech Republic} \thanks{Department of Information Management and Finance, National Chiao Tung University, Management Building 1, 1001 University Road, Hsinchu, Taiwan 30010, ROC.  }  }

\begin{document}


\maketitle
\thispagestyle{empty}


\section*{Abstract}

$K$-means clustering is one of the most widely-used partitioning algorithm in cluster analysis due to its simplicity and computational efficiency. However, $K$-means does not provide an appropriate clustering result when applying to data with non-spherically shaped clusters.\\
We propose a novel partitioning clustering algorithm based on expectiles. The cluster centers are defined as multivariate expectiles and clusters are searched via a greedy algorithm by minimizing the within cluster '$\tau$ -variance'. We suggest two schemes: fixed $\tau$ clustering, and adaptive $\tau$ clustering. Validated by simulation results, this method beats both $K$-means and spectral clustering on data with asymmetric shaped clusters, or clusters with a complicated structure, including asymmetric normal, beta, skewed $t$ and $F$ distributed clusters. Applications of adaptive $\tau$ clustering on crypto-currency (CC) market data are provided. One finds that the expectiles clusters of CC markets show the phenomena of an institutional investors dominated market. The second application is on image segmentation. compared to other center based clustering methods, the adaptive $\tau$ cluster centers of pixel data can better capture and describe the features of an image. The fixed $\tau$ clustering brings more flexibility on segmentation with a decent accuracy. All calculation can be redone via \href{https://github.com/QuantLet/KEC/tree/master/}{quantlet.com}. \\

\textbf{Keywords:}
\textit{clustering, machine learning, simulation study, parameter-tuning \\
	expectiles, partitional clustering} 
\vspace{1.0cm}

\newpage
\pagestyle{plain}
\setcounter{page}{1}    
\pagenumbering{arabic}  

\section{Introduction}
Clustering is a useful technique to discover and identify homogenous groups of data points in a given sample. As an unsupervised learning algorithm, it aims to extract information on the  underlying characteristics via dividing the data into groups that maximize common information. Obviously the information about homogeniety of groups is key in such a sample dividing mechanism. Among the simplest choice is the $K$-means clustering method described by \cite{steinhaus1956division} and \cite{hartigan1975clustering}, which adopt the Euclidean distance as neighbourhood measure, thus leading to spheres as silhouettes and means as centers of clusters. Indeed, while keeping a balance between group size and information gain, $K$-means is the most widely used partitioning algorithm due to its simplicity, efficiency in computing and easiness of interpretation. Successful applications include signal processing, image identification, customer segmentation.

The principle of a partitioning clustering algorithm is to assign data points to the nearest cluster by optimising some objective function. 
The objective function of $K$-means is the sum of within-group variance, and thus the correspondence cluster centers are the mean of each cluster.  Minimizing the objective function is equivalent to maximizing the log-likelihood function with independent Gaussian density. Although $K$-means clustering is often viewed as a "distribution free" algorithm, it is actually partitioning using equal sized spherical contour lines which can be considered as assuming independent identically distributed (i.i.d.) Gaussian clusters. 
Therefore, $K$-means approach works better for cluster in the symmetric distribution than the skewed ones.

On the other hand, when applied on skewed or asymmetric distributed data whose characteristics may not be fully captured by the first two moments, new methods are required for non-spherical cluster. To account for within-cluster skewness, \cite{hennig2019quantile} introduce the $K$-quantile clustering algorithm based on the assymmetric absoluate discrepancy. Then they linked their approach to a fixed partition model of genralized asymmetric Laplace distributions. 
This quantile discrepancy based density relies on both the quantile level $\tau$ and some additional scale/penalty parameter $\lambda$. However,  $\tau$ and $\lambda$ are assumed the same across different clusters to reduce the computation complexity. 

An analogous work on quantile based clustering is proposed in \cite{zhang2019quantile}, where they developed a model-based iterative algorithm to identify subgroups with heterogeneous slopes. In particularly, they consider clustering across multiple quantiles to capture the full picture of heterogeneity. For that accordance, how to specify the appropriate quantile level vector $\tau$ could be a problem for large dimensional data. 



This motivates us to consider a novel method, $K$-expectile clustering. This method is based on a similar idea as $K$-means but with an expectile cluster center and aims at minimizing the so-called $\tau$-variance, which is a weighted quadratic loss to take into account asymmetry. Besides being simple and fast, our algorithm can be applied on wider range of data compared with $K$-means. In particular, we consider two schemes, either with a pre-specify $\tau$ level or an adaptive $\tau$  that may vary across different dimensions or clusters, which accommodates either a fixed cluster shape or a data-driven cluster shape to capture heterogeniety. 

To better understand the basic ideas of $K$-expectile clustering, we recall some basic knowledge about tail events. Quantile regression \citep{koenker1978regression} and expectile regression \citep{newey1987asymmetric} have been suggested for displaying the whole picture of the conditional distribution of response variable on covariates, especially for data not sufficing the condition of homoskedasticity or conditional symmetry. For a random variable $X\in \mathbb{R}$ drawn from distribution $F$, a location model of $\tau$-th tail event measure with $\tau \in (0,1)$ could be defined as:
\begin{equation*}
x_{i} = \theta_{\tau}+ \varepsilon_{i}, \ i = 1,\ldots,n.
\end{equation*}
With an assumption on the $\tau$-th quantile or expectile of the cdf of $\varepsilon$ being zero, $\theta_{\tau}$ is by definition the $\tau$-th quantile or expectile of $X$ accordingly. An estimator of the location model of quantiles and expectiles can be naturally formed:
\begin{equation*}
\hat{\theta}_{\tau} = \operatorname{arg}\underset{
	\mu \in \mathbb{R}}{\operatorname{min}}\operatorname{\mathsf{E}}\left[\rho_{\tau}(X-\mu) \right],
\label{Eq2}
\end{equation*}
where the  loss function  $\rho_{\tau}(\cdot)$ is defined as:
\begin{equation*}
\rho_{\tau}(u) = |u|^{\alpha}|\tau-\operatorname{\mathbf{I}}_{\left\lbrace u\leq 0\right\rbrace }|,
\label{Eq3}
\end{equation*}
with $\alpha=1$ and $\alpha=2$ respectively. 

Although the concept of expectiles is natural analogues of quantiles, expectiles enjoy the computation efficiency over quantiles \citep{schnabel2011expectile}. In finance, the expectile might be preferred as a favorable risk measures due to its desirable properties such as coherence and elicitability (\citealp{kuan2009assessing}, \citealp{ziegel2016coherence}). Recently, the use of expectiles attracts more and more attention, such as the nonparametric expectile regression by \cite{sobotka2012geoadditive} and \cite{yang2018flexible}, the principle expectile analysis by \cite{tran2019principal}. Our proposed $K$-expectile clustering allows us to take into account tail characteristics and asymmetry when identifying homogenous groups of data, while simulation studies and applications justify its excellent performance. 

The rest of the paper is organized as follows. In section 2, we will briefly review the classical $K$-means algorithm, and then 
propose our $K$-expectile clustering in two schemes. 
In section 3, we present the simulation study that includes data from different distribution and compare the performance of $K$-expectiles clustering with other methods. Section 4 applies our method to real crypto currency market analysis and image segmentation. 
Codes of all the functions, applications and data are uploaded to \href{https://github.com/QuantLet/KEC/tree/master/}{quantlet.com}.

\section{Methodology}\label{Sec:Methodology}
\subsection{$K$-means clustering}
$K$-means clustering is rooted in signal processing, and was described by \cite{hartigan1975clustering}. Suppose the data set $X = \{X_{i}\}^{n}_{i=1}$ comes from a random sample in $\mathbb{R}^{p}$.  A clustering algorithm denoted by $Q(\cdot)$ generates $K$ subsets  $\{G_1,G_2,\ldots, G_K\}$ each with distribution $f_k(X)$. Any clustering  algorithm maps $X$ into a membership vector  $C=(c(1),c(2),\ldots, c(n))$, 
i.e. $Q(x_{i}) = c(i)$,  and $G_k =\{x_i: c(i)=k\}$,  $c(i)\in \{1,2,\ldots,K\}$.  

A clustering criterion is defined via a cost function. In $K$-means clustering, the cost is defined as the sum of squared Euclidean distance between cluster members to the cluster centroids. Indeed the centroids can be considered as location parameters for clusters. In $K$-means clustering, cluster centroids are actually cluster means and the cost function is the sum of within-cluster variance. Let $G(\cdot)$ be the $K$-means objective function,  $\Theta =  (\theta_{1}, \theta_{2}, \ldots, \theta_{K})$ be a set of cluster centroids with $\theta_{k} \in \mathbb{R}^p$,
\begin{equation*}
G(\Theta,C,X) =\operatorname{\min_{\Theta}}\sum_{k=1}^{K}\sum_{ x_{i}\in G_k}\|x_{i}-\theta_{k}\|^{2}.
\end{equation*}
Clustering is now turned into an optimisation problem and is solved via iteration. For a fixed $\Theta$, partition $c(i)$ is achieved by assigning each point to the nearest cluster centroid. 
\begin{equation*}
c(i) =\operatorname{arg}\underset{k\in\{1,2,\ldots,K\}}{\operatorname{min}}
	\|x_{i}-\theta_{k}\|^{2}.
\end{equation*}
For a fixed membership vector $C$, the centroid $\theta_{k}$ can be estimated by taking the within-cluster mean.
\begin{equation*}
\hat{\theta}_{k}= \bar{x}_{k} = \frac{1}{|G_{k}|} \sum_{ x_{i}\in G_k}x_{i}.
\end{equation*}
If $X$ are drawn i.i.d. from an unknown distribution $P$, an equally-weighted Gaussian Mixture Model can be formed by assuming the sample distribution $f(X)$ composed by a convex combination of $K$ Gaussian distributions $N_{k}(x | \mu_k,\Sigma_k)$ with expectation $\mu_k$ and variance $\Sigma_k$, 
\begin{equation*}
f(X)=  \frac{1}{K}\sum_{k=1}^{K}N_{k}(X| \mu_k,\Sigma_k),
\end{equation*}
The estimation of this model requires the EM algorithm, and the details are explained in \cite{deisenroth2020mathematics}. If let $\mu_k$ be the $k$-th cluster centrioid, and covariance matrix simply equals to $\mathcal{I}_p$, $K$-means objective function is coincide with the expectation function in EM algorithm of a Gaussian Mixture Model with equal mixture weights. By a "hard" assignment of data points to nearest cluster centroid in $K$-means algorithm as described in \cite{mackay2003information}, the computation of parameter $\Theta$ can be easily conducted by independently estimating cluster means. Simoultaneously $K$-means clustering is a distribution-free but distance based clustering technique.

\subsection{$K$-expectiles clustering}
Now consider a set of data with skewed or asymmetrically distributed clusters, i.e., cluster centroids $\Theta$ are not located on means and cluster variances are heterogeneous on different sides around centroids. As said before, it is the information measure of homogeneity that yields the clusters. A distance with cluster centroids offset from means and a distance metric which takes asymmetry into account is certainly a more flexible way of dividing groups. 

For that purpose, assume each cluster is a group of data drawn independently from a multivariate distribution $f_k(X)$, define the cluster centroid $\theta_{k}$ as the $\tau$ expectile of cluster distribution, and assign points according to expectile distances. 

More precisely, let $X \in \mathbb{R}$ be a univariate random variable with probability cumulative function $F(x)$ and finite mean $\operatorname{\mathsf{E}}|X| \leq \infty$. For a fixed $\tau \in (0, 1)$, the $\tau $-th expectile $e_{\tau}= e_{\tau} (F)$ of $X$ as proposed by \cite{newey1987asymmetric} is identified as the minimizer of the asymmetric quadratic loss

\begin{align}
&e_{\tau}(X) = \operatorname{arg}\underset{
	\mu\in\mathbb{R}}{\operatorname{min}}\operatorname{\mathsf{E}}\left[\rho_{\tau}(X-\mu) \right],  \label{eq1}\\
&\rho_{\tau}(X-\mu) = \tau(X-\mu)_{+}^{2} + (1-\tau)(\mu-X)_{+}^{2}, \label{eq2}
\end{align}
 where $(x)_{+}=\operatorname{max}(x,0)$. it is worth noting here that the expectile location estimator can be interpreted as a maximum likelyhood estimator of a normal distributed sample with an unequal weight placed on positive and negative disterbances, showed in \cite{aigner1976estimation}. 
 
For $X\in \mathbb{R}^{p}$, define  $(X)_{+} = ((X_{1})_{+}, \dots , (X_{p})_{+})^{\top}$, then the multivariate expectile $e_{\tau}(X)\in \mathbb{R}^{p}$ is the solution to the optimization problem:
\begin{equation}\label{eq3}
e_{\tau}(X) = \operatorname{arg}\underset{
	\mu\in\mathbb{R}^p}{\operatorname{min}}\operatorname{\mathsf{E}}\left[ \tau\|(X-\mu)_{+}\|^{2} + (1-\tau)\|(\mu-X)_{+}\|^{2} \right].
\end{equation}

Here the dependence is taken into account by using the norm. The construction of the multivariate expectiles are related to the dependence structure of each components. The choice of dependence modelling may differ according to the practical goal. For simplification reasons, we only elaborate the case when the dependence structure is ignored. The multivariate expectile $e_{\tau} (X)$ now consists of marginal univariate expectiles,
\begin{align*}
e_{\tau}(X) &= \operatorname{arg}\underset{
	\mu\in\mathbb{R}^{p}}{\operatorname{min}}\operatorname{\mathsf{E}}\left[\tau\{\sum_{j=1}^{p}\|x_{\bigdot j}-\mu_{j}\|_{+}^{2}\}+(1-\tau)\{\sum_{j =1}^{p}\|\mu_{j}-x_{\bigdot j}\|_{+}^{2}\} \right] \\
&=\operatorname{arg}\underset{
	\mu\in\mathbb{R}^{p}}{\operatorname{min}}\operatorname{\mathsf{E}}\left[\sum_{j =1}^{p}\{{\tau}\|x_{\bigdot j}-\mu_{j}\|_{+}^{2}+(1-\tau)\|\mu_{j}-x_{\bigdot j}\|_{+}^{2}\} \right] \\
&=(e_{\tau}(x_{\bigdot 1}),\ldots,e_{\tau}(x_{\bigdot p}))^{\top},
\end{align*}
where $x_{\bigdot j}$ denotes the $j$-th coloum of matrix $X$ \citep{maume2016multivariate}.

The flexibility and power of expectiles in $p$ dimension comes from looking at $\tau = (\tau_{1}, \tau_{2}, \ldots, \tau_{p})^\top\in \mathbb{R}^p$, making the $\tau$ level variable over dimensions. Thus, we obtain
\begin{align*}
e_{\tau}(X)&=\operatorname{arg}\underset{
	\mu\in\mathbb{R}^{p}}{\operatorname{min}}\operatorname{\mathsf{E}}\left[\sum_{j=1}^
{p}\{{\tau_{j}}\|x_{\bigdot j}-\mu_{j}\|_{+}^{2}+(1-\tau_{j})\|\mu_{j}-x_{\bigdot j}\|_{+}^{2}\} \right] \\
&=(e_{\tau_1}(x_{\bigdot 1}),\ldots,e_{\tau_p}(x_{\bigdot p}))^{\top}.
\end{align*}
The empirical version reads as:
\begin{equation}\label{eq4}
\hat{e}_{\tau,n}(X)=\operatorname{arg}\underset{
	\mu\in\mathbb{R}^{p}}{\operatorname{min}}\frac{1}{n}\sum_{i}^{n}\sum_{j=1}^
{p}\{{\tau_{j}}(x_{i j}-\mu_{j})_{+}^{2}+(1-\tau_{j})(\mu_{j}-x_{i j})_{+}^{2}\} \\.
\end{equation}

The idea is now to fix the cluster centroids at the empirical expectile of the $k$-th cluster, i.e. $\theta_{k} = e_{\tau,k}(x_i\in G_k)$, and consider an asymetrically weighted distance function with $L_{2}$ norm. For an observation $x \in \mathbb{R}^{p}$, define $\tau$-distance
\begin{equation}\label{eq5}
d(x,\tau,\theta) = \left\lbrace \tau +(1-2\tau)  \operatorname{\mathbf{I}}_{\left\lbrace x<\theta\right\rbrace } \right\rbrace \|x-\theta\|^{2},
\end{equation}
which coincides with loss function (\ref{eq2}).

Based on the concept of $\tau$-distance, instead of specifying an asymmetric form of distribution $f(X)$ or $f_k(X)$, we can form a $K$-expectile objective function by defining an asymmetric $\tau$-variance as described in \cite{tran2019principal}, 
\begin{equation*}
v(x,\tau,\theta) = \frac{1}{n}\sum_{i}^{n}\sum_{j=1}^
{p}\{{\tau_{j}}(x_{i j}-\theta_{j})_{+}^{2}+(1-\tau_{j})(\theta_{j}-x_{i j})_{+}^{2}\} 
\end{equation*}
which yields to an axis-aligned elipsoid unit ball. To include covariance or correlation, usually a matrix form of multivariate expctile will be considered. By introducing a $p\times p$ symmetric matrix $\Sigma$, one can form a score function as described in \cite{maume2016multivariate},
\begin{equation}\label{eq18}
e_{\tau}^{\Sigma}(X) \in \operatorname{arg}\underset{
	\mu\in\mathbb{R}^{p}}{\operatorname{min}}\operatorname{\mathsf{E}}\left[\tau(x-\mu)_{+}^{\top}\Sigma(x-\mu)_{+}+(1-\tau)(x-\mu)_{+}^{\top}\Sigma(x-\mu)_{+}\right].
\end{equation}
However, we will not include this case in this paper. We will leave this problem for further works. 

\begin{figure}[ht]
		\makebox[\textwidth][c]{\includegraphics[scale=0.6,angle=0]{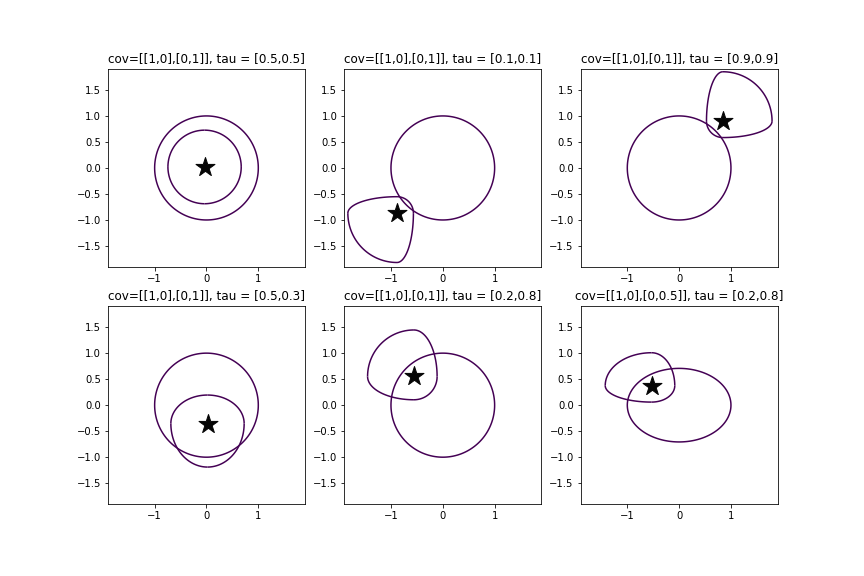}}
		\caption{Contour lines of unit balls of various $\tau$ variances in comparison to unit balls of symmetric variance}\includegraphics[scale=0.05]{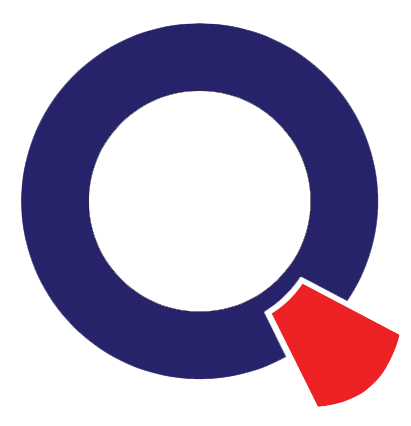}\href{https://github.com/QuantLet/KEC/tree/master/KEC_cluster\%20shapes}{KEC\_cluster shapes}
		\label{Fig:various}
\end{figure}

\begin{figure}[ht]
		\makebox[\textwidth][c]{\includegraphics[scale=0.4,angle=0]{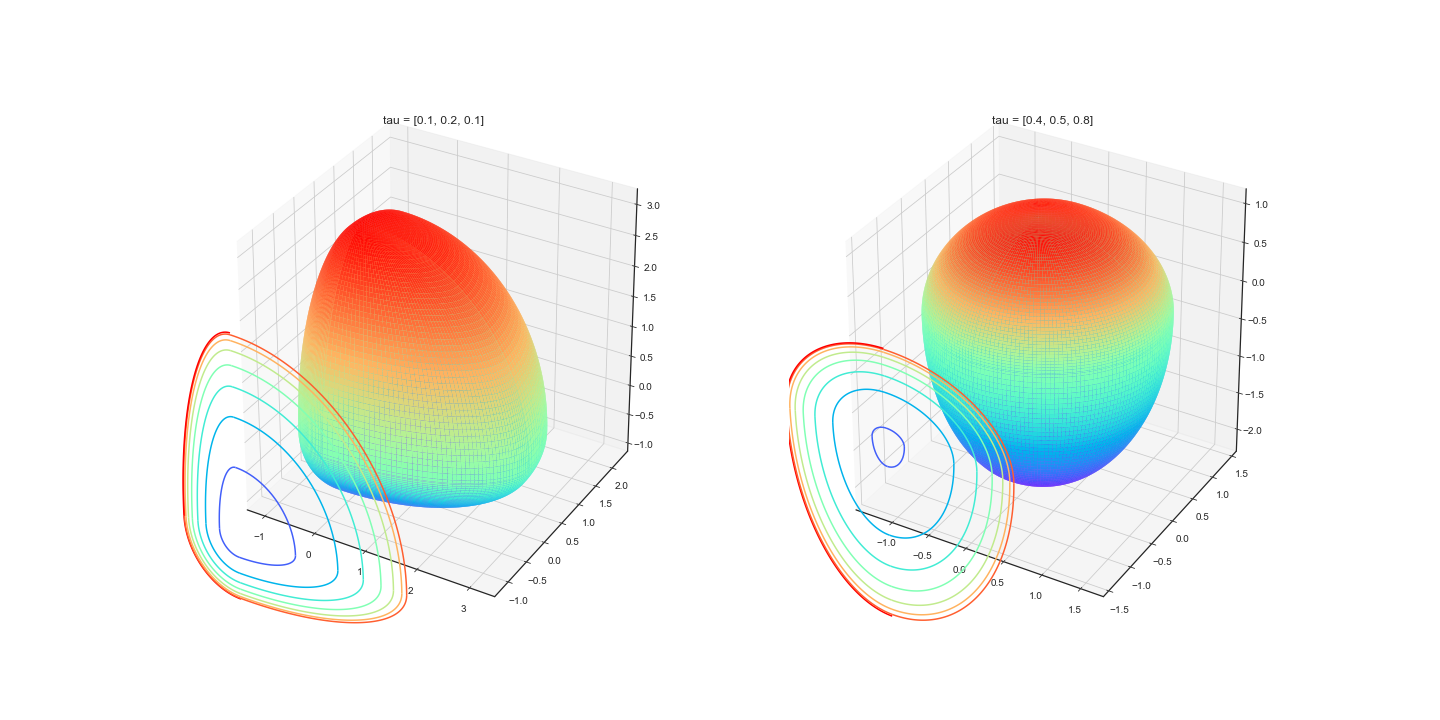}}
		\caption{3D contour surface of unit balls of various $\tau$ variances. Left: $\tau = [0.1, 0.2, 0.1]$. Right: $\tau = [0.4, 0.5, 0.8]$.}\includegraphics[scale=0.05]{qletlogo_tr.png}\href{https://github.com/QuantLet/KEC/tree/master/KEC_cluster\%20shapes}{KEC\_cluster shapes}
		\label{Fig:3D}
\end{figure}

In Figure \ref{Fig:various}, the contour lines of unit circles of bivariate $\tau$-variance with various $\tau$ levels on each axis are shown along with unit circles of a symmetric variance in the back. The covariance matrix is the inverse matrix of $\Sigma$ in function \ref{eq18}. The last sub-plot shows the unit circles with different scales on two axis. These are equivalent to the contour lines of independent bivariate asymmetric normal distributions in comparison with the contour lines of independent normal distributions. Figure \ref{Fig:3D} shows the 3D contour surface of $\tau$-variance unit ball with different $\tau$-levels on each dimension, or the 3D cluster shapes.

\subsection{Fixed $\tau$ clustering}
With distance (\ref{eq5}) and a pre-specified $\tau\in \mathbb{R}^p$ vector, define the objective function
\begin{align}
G^{Fixed}(\tau,\Theta,C,X)&=\sum_{k=1}^{K}\sum_{x_i\in G_k}\sum_{j=1}^{p}d(x_{i j},\tau_{j},\theta_{k})  \label{eq6}\\
&=\sum_{k=1}^{K} \sum_{x_i\in G_k}\sum_{j=1}^{p} \left\lbrace\tau_{j}+(1-2\tau_{j})  \operatorname{\mathbf{I}}_{\left\lbrace x_{i j}<\theta_{k,j}\right\rbrace } \right\rbrace (x_{i j}-\theta_{k,j})^{2}, \label{eq7}
\end{align}
which aims to detect expectile-specified clusters by minimizing the sum of within-cluster $\tau$-variance. 

For known $(\tau,C)$, cluster centroids $\Theta$ are found by:
\begin{align}
\hat{\theta}_{k}&= \operatorname{arg}\underset{\mu\in\mathbb{R}^p}{\operatorname{min}}\;\sum_{x_i\in G_k}\sum_{j=1}^{p} \left\lbrace\tau_{j}+(1-2\tau_{j})  \operatorname{\mathbf{I}}_{\left\lbrace x_{i j}<\mu_{j}\right\rbrace } \right\rbrace (x_{i j}-\mu_{j})^{2} \label{eq8}\\
&= \operatorname{arg}\underset{
	\mu\in\mathbb{R}^p} {\operatorname{min}}\; \sum_{x_i \in G_k}\sum_{j=1}^{p} w_{ij}(\tau_j) (x_{ij} - \mu_{j} )^{2} \label{eq9}
\end{align}

where $w(\tau)$ is a weight function which is related to $\mu(\tau)$, the location parameter at the given $\tau$ level.
\begin{equation}\label{eq10}
w_{ij}(\tau_j) =\begin{cases}
	\tau_j       & \text{if} \ x_{ij} \leq \mu_{j}(\tau_j)  \\
	1 - \tau_j  & \text{if} \ x_{ij} > \mu_{j}(\tau_j).
\end{cases}
\end{equation}

This implicit dependence of $w$ on $\mu(\tau)$ leads to the application of the  Least Absolute Square Estimator (LAWS), a version of the Stochastic Gradient Algorithm. For a fixed $\mu_{j}(\tau_j)$, the weight $w_{ij}(\tau_j)$ in equation (\ref{eq10}) is calculated, therefore a closed form solution of $\mu_{j}(\tau_j)$ can be expressed as
\begin{equation}\label{eq11}
\mu_{j}(\tau_j) = \frac{\tau_j\sum_{i \in \mathcal{I}_{\tau_j}^{+}}x_{ij} + (1-\tau_j)\sum_{i \in \mathcal{I}_{\tau_j}^{-}}x_{ij}}{\tau_j n^{+} + (1-\tau_j)n^{-}}
\end{equation}
where
\begin{align*}
&\mathcal{I}_{\tau}^{+} = \{i \in \{1, \ldots, n\}: w_{ij}  = \tau_j, c(i)=k\}\\
&\mathcal {I}_{\tau}^{-} = \{i \in \{1, \ldots, n\}: w_{ij}  = 1-\tau_j, c(i)=k\}\\
&n^{+}=|\mathcal I_\tau^+|   \qquad    n^{-}=|\mathcal I_\tau^-|.
\end{align*}
Cluter centroids can be estimated by iteratively repeating the two steps until the location of $\mu_{j}(\tau_j)$ does not change, see:\\
\begin{algorithm}[htb] 
	\caption{ LAWS} 
	\label{Algo:LAWS} 
	\begin{algorithmic}[1] 
		\Require 
		The set of points in cluster $G_k$; 
		The vector of parameter, $\tau$; 
		\Ensure 
		Estimated cluster centroids, $\Theta$
		\State Initialize $\mu^0_{j}(\tau_j) $ as mean of $j$-th column of $x_i \in G_k$
		\label{code4} 
		\Repeat 
		\State Assign weight $w_{ij}^{t+1}(\tau_j)$ to each point $x_{ij}$ based on $\mu^t(\tau)$
		\State Update $\mu^{t+1}(\tau)$ according to equation (\ref{eq11}) with input $w^{t+1}(\tau)$ 
		\Until{$d\{\mu(\tau)^{t},\mu(\tau)^{t-1}\} = 0$}  
	\end{algorithmic} 
\end{algorithm}

The $K$-expectiles clustering algorithm now read as follows:\\
\begin{algorithm}[htb] 
	\caption{ Fixed $\tau$ clustering} 
	\label{Algo:Fixed} 
	\begin{algorithmic}[1] 
		\Require 
		Data, $X$; 
		Vector parameter, $\tau$; 
		\# of clusters, $K$; 
		\Ensure 
		Cluster membership vector, $C$; 
		Estimated cluster centroids, $\Theta$
		\State Initialize centroids  $\Theta^0 = \Theta_{k-means}$
		\label{code1} 
		\Repeat 
		\State Calculate cluster membership $c(i)^{t+1} =\operatorname{arg}\underset{k\in\{1,2,\ldots,K\}}{\operatorname{min}}
		 \sum_{j=1}^{p} \sum_{i:c(i)^{t}=k}\left\lbrace\tau_{j}+(1-2\tau_{j})  \operatorname{\mathbf{I}}_{\left\lbrace x_{i j}<\theta^t_{k,j}\right\rbrace } \right\rbrace \|x_{i j}-\theta^t_{k,j}\|^{2}$
		\State Update $\theta^{t+1}_{k}$ by applying Algorithm (\ref{Algo:LAWS}) with input $c(i)^{t+1}$
		\Until{$d\{\theta^{t},\theta^{t-1}\}\leq threshold$}  
	\end{algorithmic} 
\end{algorithm}

\subsection{Adaptive $\tau$ clustering}
\begin{figure}[ht]
		\makebox[\textwidth][c]{\includegraphics[scale=0.4,angle=0]{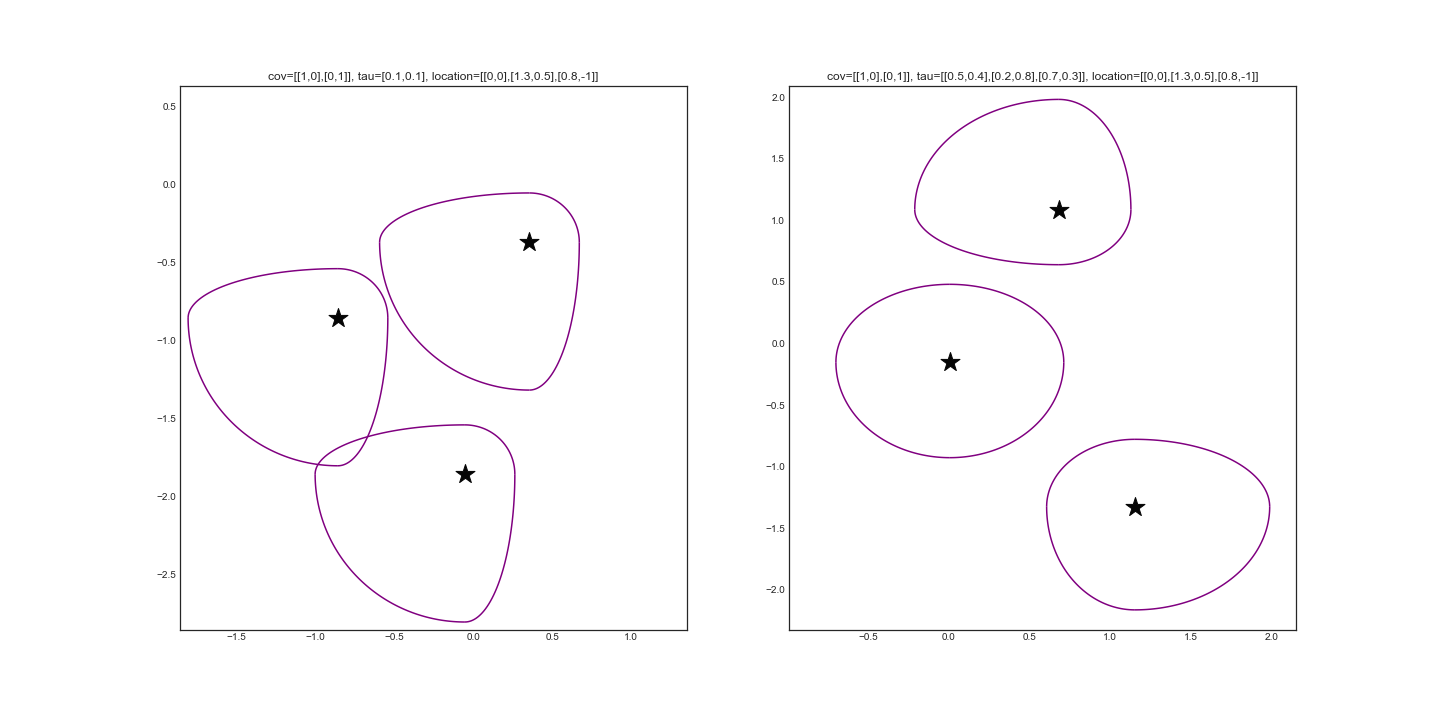}}
		\caption{Different senarios regarding to cluster shapes}\includegraphics[scale=0.05]{qletlogo_tr.png}\href{https://github.com/QuantLet/KEC/tree/master/KEC_cluster\%20shapes}{KEC\_cluster shapes}
		\label{Fig:clusters}
\end{figure}

In the last section, clustering with fixed cluster shapes by pre-specifying the $\tau$ vector has been discussed, and this senario is shown in Sub-plot $1$ of Figure \ref{Fig:clusters}. In comparison, regarding the issue of clusters with different shapes as shown in Sub-plot $2$ of Figure \ref{Fig:clusters}, we present the result of a fully adaptive algorithm, both for different dimensions and for different clusters. Without pre-defined $\tau$, we now assume $\tau \in (\mathbb{R}^p)^K$ is a $(p\times K)$ matrix. We optimize the following cluster objective function with respect to  $\tau$ as well. 

\begin{align}
G^{Adaptive}(\tau,\Theta,C,X)&=\sum_{k=1}^{K}\sum_{x_i \in G_k}\sum_{j=1}^{p}d(x_{i j},\tau_{k,j},\theta_{k,j}) \label{eq12}\\
&=\sum_{k=1}^{K} \sum_{x_i \in G_k}\sum_{j=1}^{p} \left\lbrace\tau_{k,j}+(1-2\tau_{k,j})  \operatorname{\mathbf{I}}_{\left\lbrace x_{i j}<\theta_{k,j}\right\rbrace } \right\rbrace (x_{i j}-\theta_{k,j})^{2}, \label{eq13}
\end{align}

For given $(\Theta,C)$, to optimize $\tau$, require
\begin{equation}\label{eq14}
\hat{\tau}_{k}= \operatorname{arg}\underset{\tau\in(\mathbb{R}^p)^K}{\operatorname{min}}\;
G^{Adaptive}(\tau,\Theta,C,X).
\end{equation}
By taking first order condition, we get the unique solution:
\begin{equation}\label{eq15}
\tau_{k,j} = \frac {\gamma_{k,j}}{1 + \gamma_{k,j}},
\end{equation}
where 
\begin{equation*}
\gamma_{k,j} = \frac{n^-\sum_{i \in \mathcal{I}_{\tau}^+}\theta_{k,j}-x_{ij}}{n^+\sum_{i \in \mathcal{I}_{\tau}^+}x_{ij}-\theta_{k,j}}.
\end{equation*}

Then the clustering algorithm for adaptive $\tau$ can be described as:
The clustering algorithm can be implemented as follows:\\
\begin{algorithm}[htbp] 
	\caption{ Adaptive $\tau$ clustering} 
	\label{Algo:Adaptive} 
	\begin{algorithmic}[1] 
		\Require 
		Data, $X$; 
		\# of clusters, $K$; 
		\Ensure 
		Cluster membership vector, $C$; 
		Estimated cluster centroids, $\Theta$
		\State Initialize centroids  $\Theta^0 = \Theta_{k-means}$;
		$\tau^{0}_{k,j} = 0.5$
		\label{code3} 
		\Repeat 
		\State Calculate cluster membership $c(i)^{t+1} =\operatorname{arg}\underset{k\in\{1,2,\ldots,K\}}{\operatorname{min}}
		\sum_{j=1}^{p} \sum_{i:c(i)^{t}=k}\left\lbrace\tau^t_{k,j}+(1-2\tau^t_{k,j})  \operatorname{\mathbf{I}}_{\left\lbrace x_{i j}<\theta^t_{k,j}\right\rbrace } \right\rbrace \|x_{i j}-\theta^t_{k,j}\|^{2}$
		\State Update $\theta^{t+1}_{k}$ by applying Algorithm (\ref{Algo:LAWS}) with input $\tau^t_{k}$
		\State Update $\tau^{t+1}_{k}$ according to equation (\ref{eq15}) with input $\theta^{t+1}_{k}$
		\Until{$d\{\theta^{t},\theta^{t-1}\}\leq threshold$}  
	\end{algorithmic} 
\end{algorithm}

\newpage
\section{Simulation}\label{Sec:Simulation}

To evaluate the performance of $K$-expectiles clustering, we design four simulated samples with $K$ clusters. Let $k = 1,\ldots, K$, each cluster represented by $G_k$ is an i.i.d. random sample drawn from a $p$-variate distribution in the size of $(n_1, n_2, \ldots, n_k)$. Each component of the multivariate distribution is assumed to be independent. Data set can be written as $X = (G_1, G_2,\ldots,G_k)$. Scale, location and skewness of the distribution can cause the overlapping of multiple clusters which in turn influece the cluster shapes and within-cluster data density, thus hinder the accuracy of grouping results. The simulated samples are designed to  reserve some extend of overlap while ensure certain discrimination between clusters, in order to achieve the purpose of evaluating the robustness of the algorithms.
 
\begin{description}
	\item[Sample 1:] In the first sample we generate $K$ multivariate Gaussian clusters with unit variance and different location parameters. $G_{k} \sim N(\mu_k, \mathcal{I}_{p})$, where $\mu_{1}$ is a $p$-dimensional integer vector whose elements are randomly generated in interval $(1,10)$, and then shift the location of other clusters by $\mu_{k}= \mu_{1} + 2k$. Clusters are in the same size of $n_k = n/k$.
	
	\item[Sample 2:] To include some asymmetry on the basis of Sample 1, the second sample is designed as a mixture of $K$ asymmetric normal distributions. Each cluster $G_{k}$ is considered as a $p$-dimensional i.i.d. sub-sample, where $G_{k} = (W_{1}, W_{2},\ldots,W_{p})^{\top}$.
    The probability density function of $W_{j}$ can be expressed as following, with $\mu_j (j= 1, 2, \ldots, p)$ as location parameters, and $\sigma_{l}$, $\sigma_{r}$ as standard deviation of two sides around $\mu$,
    \begin{equation*}
    p(W_{j}\mid \theta) = \prod_{j=1}^{p} \sqrt{\frac{2}{\pi}}    \frac{1}{\sigma_{l_{j}}+\sigma_{r_{j}}}  \left\{
    \begin{aligned}
    	\operatorname{exp}\left\lbrace  - \frac{(W_{j}-\mu_{j})^{2}}{2\sigma^{2}_{l_{j}}}\right\rbrace  &      & {W_{j} < \mu_{j}} \\
    	\operatorname{exp}\left\lbrace- \frac{(W_{j}-\mu_{j})^{2}}{2\sigma^{2}_{l_{j}}}\right\rbrace&      & {W_{j} \geq \mu_{j}} \\
    \end{aligned}
    \right.,
   \end{equation*}
  Now let $e_{\tau}$ ($\tau$-expectile of the variable) be the location parameter, and $\sigma_{j}$ be the overall standard deviation of the variable, the density function of asymmetric normal distribution can be rewritten as:
   \begin{equation*}
	p(Z_{j}\mid e_{\tau}, \sigma_{j}, \tau) =   \left\{
	\begin{aligned}
	\prod_{j=1}^{p} \sqrt{\frac{1}{2\pi(\tau^{-\frac{1}{2}}\sigma_{j})}}\frac{2\sqrt{1-\tau}}{\sqrt{1-\tau}+\sqrt{\tau}}\operatorname{exp}\left\lbrace- \frac{(Z_{j}-e_{\tau})^{2}}{2(\tau^{-\frac{1}{2}}\sigma_{j})^{2}} \right\rbrace&      & {Z_{j} < e_{\tau}} \\
	\prod_{j=1}^{p} \sqrt{\frac{1}{2\pi((1-\tau)^{-\frac{1}{2}}\sigma_{j})}}\frac{2\sqrt{\tau}}{\sqrt{1-\tau}+\sqrt{\tau}}\operatorname{exp}\left\lbrace - \frac{(Z_{j}-e_{\tau})^{2}}{2((1-\tau)^{-\frac{1}{2}}\sigma_{j})^{2}}\right\rbrace&      & {Z_{j} \geq e_{\tau}}\\
	\end{aligned}
	\right.,
	\end{equation*}
   which means the asymmetric normally distributed variable $W_j (j= 1, 2, \ldots, p)$ can be converted from univariate Gaussian distributed variables $Z_j$ according to the formula:
   \begin{equation*}
   	W_j^k=   \left\{
   	\begin{aligned}
   		\frac{2\sqrt{\tau^k_{j}}}{\sqrt{1-\tau^k_{j}}+\sqrt{\tau^k_{j}}} \cdot \frac{1}{\sqrt{1-\tau^k_{j}}}\cdot Z_{j}+e_{\tau^k_{j}}&      & {Z_{j}^k < 0} \\
   		\frac{2\sqrt{1-\tau^k_{j}}}{\sqrt{1-\tau^k_{j}}+\sqrt{\tau^k_{j}}}  \cdot \frac{1}{\sqrt{\tau^k_{j}}}\cdot Z_{j}+e_{\tau^k_{j}}&      & {Z_{j}^k \geq 0}\\
   	\end{aligned}
   	\right.,
   \end{equation*}
   in our sample, each $Z_{j} ^{k}\sim N(0, 25)$. Parameter $\tau^{k}_{j}$ is given by using random generator with interval $[0.1, 0.9]$, and location parameter $e_{\tau^{1}_{j}}$ is randomly generated in $(0,10)$$k$-th cluster, then the location of $k$-th cluster can be shifted by $e_{\tau^{k}_{j}} = e_{\tau^{1}_{j}} + 7(-1)^{j}(j-1)$.
	
    \item[Sample 3:] In the third sample we want to test the algorithm on skewed but not leptokurtic clusters, namely $Beta$-distributed clusters. For variables in cluster$\{k=2c+1, c\in\mathbb{Z}\}$, $W_j^i \sim \mathit{Beta}(a_j,b_j)$, $(j= 1, 3, \ldots, p-1)$, and in cluster  $\{k=2c, c \in \mathbb{Z}\}$, $W_j^i \sim \mathit{Beta}(b_j,a_j)$, $(j= 2, 4, \ldots, p)$. We generate parameter $a_j$ randomly from interval $(1,10)$ and $b_j$ from interval $(10,20)$, again $K=3$.

    \item[Sample 4:] For the last sample, skewed and leptokurtic clusters are being considered.
    We set $2$ different scenarios:
    \begin{itemize}
    	 \item $K$ skewed generalized $t$-distributed samples. We first generated a random sample with dimension $p=2$, parameters $df = [10, 10, 10], nc = [3, -1.5, 2.5]$, location randomly fluctuated with the difference of 0.5 around $ [[0,2],[1,0],[0.5,1]]$, $scale = 0.5$. And generate data repeatedly until $p$ reaches $10$ and $50$.
    	 
    	\item $K$ multivariate $F$-distributed clusters. For variables in the first cluster, $W_{j}^{1} \sim \mathit{F}(a_j,a_j)+1$, and when $j= 1, 3, \ldots, p-1$; $W_{j}^{1} \sim \mathit{F}(b_j,b_j)+1$,  when  $j= 2, 4, \ldots, p$, where $a_{j}$  and  $b_{j}$ are integers randomly selected from interval $(51,60)$ and $(21,30)$. In the second cluster,  $W_{j}^{2}\sim \mathit{F}(b_j,b_j)$, $j= 1, 3, \ldots, p-1$, $W_{j}^{2} \sim \mathit{F}(a_j,a_j)$, $j= 2, 4, \ldots, p$, where $a_{j}$ and $b_{j}$ are integers randomly selected from interval $(5,15)$ and $(25,35)$. In the third cluster, $W_{j}^{3} \sim \mathit{F}(a_j,b_j)$, $j=1,3,\ldots,p-1$, and $W_{j}^{3} \sim \mathit{F}(b_j,a_j)$, $j=2,4,\ldots,p$, where $a_{j}$ and $b_{j}$ are integers randomly selected from interval $(15,25)$ and $(60,70)$.

    \end{itemize}
\end{description}
For each of the first three samples, we evaluate combinations of $p = 50, 100, 500, n =  300, 1500$. For the last sample, $p = 2, 10, 50$. 

For each simulation setting, we re-generate the data $50$ times and test the algorithms each round, and take the average of the Adjusted Rand Index (ARI) of the yielded classification compared with the true cluster membership. Rand Index measures the pair-wised agreement between data clustering.  When it is djusted for the chance grouping of elements, this is the Adjusted Rand Index. Given two partitions $X={X_1,X_2, \ldots, X_r}$ , $Y={Y_1,Y_2, \ldots, Y_s}$, and the contingency table,

\begin{equation*}
\begin{array}{c|cccc|c}{{} \atop X}\!\diagdown \!^{Y}&Y_{1}&Y_{2}&\cdots &Y_{s}&{\text{sums}}\\\hline X_{1}&n_{11}&n_{12}&\cdots &n_{1s}&a_{1}\\X_{2}&n_{21}&n_{22}&\cdots &n_{2s}&a_{2}\\\vdots &\vdots &\vdots &\ddots &\vdots &\vdots \\X_{r}&n_{r1}&n_{r2}&\cdots &n_{rs}&a_{r}\\\hline {\text{sums}}&b_{1}&b_{2}&\cdots &b_{s}&\end{array},
\end{equation*}
the Ajusted Rand Index is defined as:
\begin{equation*}
 ARI={\frac {\left.\sum _{ij}{\binom {n_{ij}}{2}}-\left[\sum _{i}{\binom {a_{i}}{2}}\sum _{j}{\binom {b_{j}}{2}}\right]\right/{\binom {n}{2}}}{\left.{\frac {1}{2}}\left[\sum _{i}{\binom {a_{i}}{2}}+\sum _{j}{\binom {b_{j}}{2}}\right]-\left[\sum _{i}{\binom {a_{i}}{2}}\sum _{j}{\binom {b_{j}}{2}}\right]\right/{\binom {n}{2}}}}.
\end{equation*}

Other distance based clustering algorithms such as $K$-means denoted by \emph{$K$-means}, spectral clustering \citep{shi2000normalized} denoted by \emph{spectral}, Ward hierarchical clustering \citep{ward1963hierarchical} denoted by \emph{h-ward}, and Quantile based clustering \citep{hennig2019quantile} are comparing with $K$-expectile clustering with adaptive $\tau$. Note that Quantile based clustering algorithm allows quantile level (skewness parameter) $\tau$ to change variable-wisedly and introduces a scale/penalty parameter. \emph{CU, CS, VU, VS} stands for the four modes of Quantile based clustering, corresponding to Common skewness parameter and Unscaled variables, Common skewness parameter and Scaled variable-wise, Variable-wise skewness parameter and Unscaled variables, Variable-wise skewness parameter and Scaled variable-wise. Results are shown in Apendix from Table \ref{tab:sim1}  to table \ref{tab:sim4-2}, where the demonstrated values are the $100$ times of ARI.

 The cluster algorithm shows a higher ARI score when the sample sizes are larger and the dimensionalities are higher. Spectral clustering sometimes does not work appropriate on data with outliers which lead to a not fully connected graph. This scenario can be easily occured in a highly skewed sample or sample with large dimensionality. 

From Table \ref{tab:sim1} we can conclude that $K$-expectile, as an algorithm that generalize $K$-means, works as good as but sometimes even better than $K$-means on spherical clusters. Meanwhile it is better than all the other clustering algorithms, incluing Ward hierarchical clustering, spectral clustering, and quantile based clustering.

For asymmetric normal distributed clusters, as Table \ref{tab:sim2} shows, $K$-expectile outperforms all the listed algorithms. Since the contour lines of the real distribution of the data correspond to the assumption of $K$-expectile cluster shapes, $K$-expectile yields a significantly better result than other algorithms.

For more general skewed distributed clusters as demonstrated in Table \ref{tab:sim2}, \ref{tab:sim3}, \ref{tab:sim4-1} and \ref{tab:sim4-2}, $K$-expectile still has a robust and outstanding performance. It always performs better than $K$-means, only except for some special cases of low-dimensional beta distributed clusters, which may due to the almost sphere cluster shapes and the non-leptokurtic density. On this kind of samples, Ward hierarchical clustering has a moderate but non-robust performance, especially when the dimensionality goes higher. Quantile based clustering, in another hand, has a comparable performance on skewed data with $K$-expectile. Under non-leptokurtic scenario, 
$K$-expectile performs better, while working on leptokurtic clusters, among the four algorithms of Quantile based clustering, some has superior results but others can be inferior. Criterion of selection among the four algorithms needs to be further studied.


\newpage
\section{Application}\label{Sec:Application}
\subsection{Application of adaptive $\tau$ clustering on CRIX and VCRIX data}
An application based on the CRIX and VCRIX data is presented in this section. CRIX (CRyptocurrency IndeX) developed by \cite{trimborn2018crix} provides a CC market price index weighted by market capitalization with a dynamic changing number of constituents of representative cryptos. The mechanism of selecting CRIX constituents with Akaike Informstion Creterion is introduced in the mentioned paper. VCRIX, developed by \cite{kim2019vcrix} is a volatility index built on CRIX which offers a forecast for the mean annualized volatility of the next 30 days, re-estimated daily by  using Heterogeneous Auto-Regressive (HAR) model.

The data are downloaded from \href{https://thecrix.de}{thecrix.de}, consists of two time series, CRIX and VCRIX, collected daily from 2017-01-02 to 2021-02-09, in total 1497 observations in two dimensions. Here we scaled the data by dividing each varaible by their standard deviations to ensure the data has equal variance. The descriptive statistics and density plots of the two variables are listed as following.

  \begin{table}[ht]
	
	\begin{center}
		{\footnotesize
			\begin{tabular}{l|cccccccccc}
				\hline \hline
				& Min.    & 1st Qu.   & Median   & Mean   & 3rd Qu.   & Max   & Skewness   & Kurtosis  & JB statistic   \\
				\hline
				CRIX   & 0.080 & 0.621 & 1.056 & 1.246 & 1.443 & 7.257 & 2.450 & 10.988 & 5481.283 \\
				VCRIX    & 0.801 & 1.814 & 2.225 & 2.458 & 2.884 & 6.565 & 1.370 & 5.360 & 816.166  \\
				\hline \hline
		\end{tabular}}
	\end{center}
	\caption{Descriptive statistics of location and dispersion for
		1497 scaled data for the period from January 02, 2017
		to February 09, 2021. }
	\label{Tab:DescripStatsCRIX-VCRIX}
\end{table}

\begin{figure}[ht]
	\begin{center}
		\includegraphics[scale=0.6,angle=0]{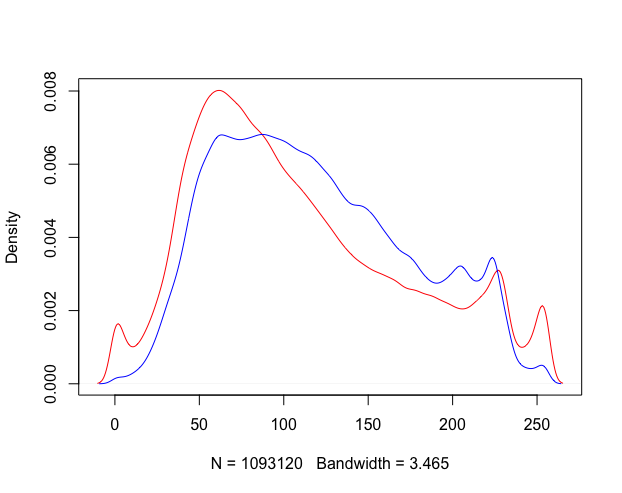}
		\caption{Variable densities. The red line and the blue line are the result of kernel density estimation of scaled CRIX and VCRIX.}\includegraphics[scale=0.05]{qletlogo_tr.png}\href{https://github.com/QuantLet/KEC/tree/master/KEC_applications}{KEC\_application}
		\label{Fig:variable densities}
	\end{center}
\end{figure}

Looking at Table \ref{Tab:DescripStatsCRIX-VCRIX}, it is evident that neither of the two variables are normally distributed and both of them are skewed. This fact can be seen in Figure \ref{Fig:variable densities} as well, due to the longer right tail of the densities of both variables. From the plot of marginal distribution one might suspect several clusters exist.

\begin{figure}[htbp]
	\begin{center}
		\includegraphics[scale=0.4,angle=0]{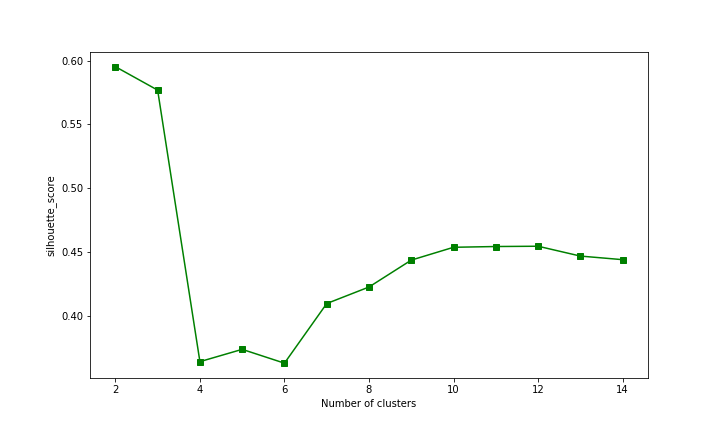}
		\caption{Sillouette score of $K$-expectiles clustering results with different number of clusters }\includegraphics[scale=0.05]{qletlogo_tr.png}\href{https://github.com/QuantLet/KEC/tree/master/KEC_applications}{KEC\_application}
		\label{Fig:silouette}
	\end{center}
\end{figure}

\begin{figure}[htbp]
	\begin{center}
		\includegraphics[scale=0.4,angle=0]{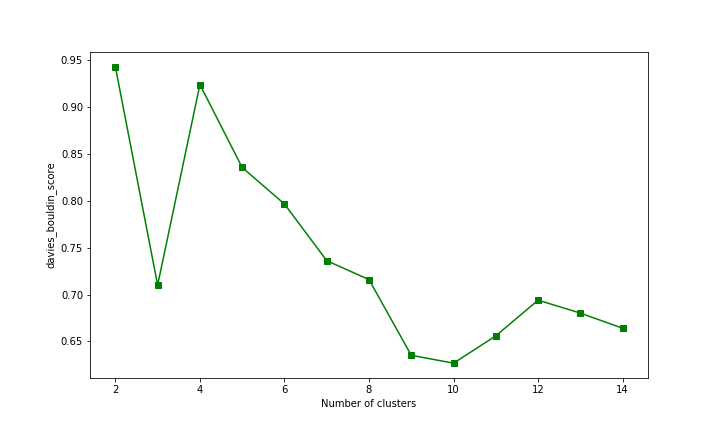}
		\caption{Davies-Bouldin scoreof $K$-expectiles clustering results with different number of clusters }\includegraphics[scale=0.05]{qletlogo_tr.png}\href{https://github.com/QuantLet/KEC/tree/master/KEC_applications}{KEC\_application}
		\label{Fig:db}
	\end{center}
\end{figure}

\begin{figure}[htbp]
	\begin{center}
		\includegraphics[scale=0.4,angle=0]{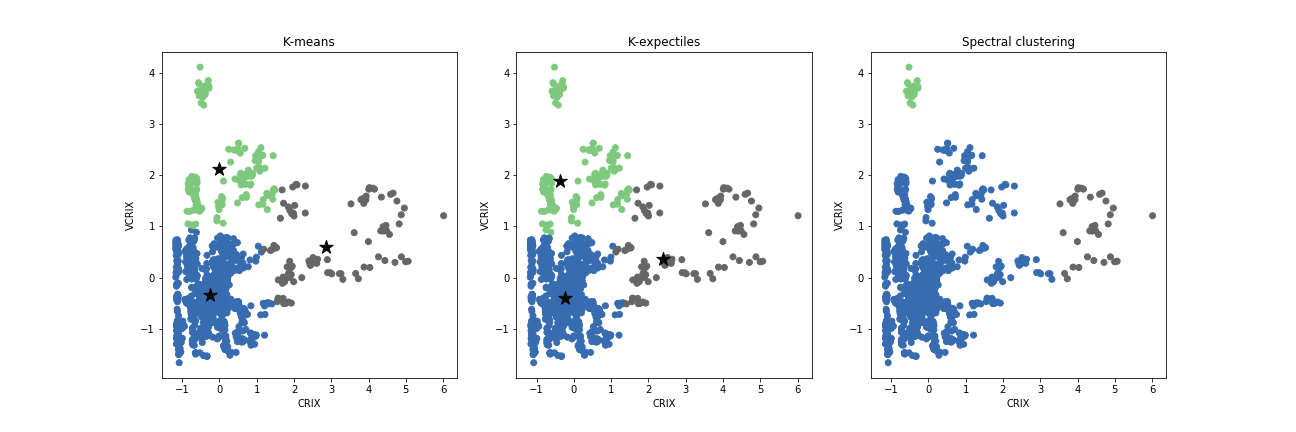}
		\caption{Clustering results of $K$-means, $K$-expectiles and Spectral clustering. The two variables are plot along x-axis and y-axis. Clusters are shown in different colours, whereas cluster centroids are shown by stars. }\includegraphics[scale=0.05]{qletlogo_tr.png}\href{https://github.com/QuantLet/KEC/tree/master/KEC_applications}{KEC\_application}
		\label{Fig:me}
	\end{center}
\end{figure}

Results of $K$-means clustering, $K$-expectile clustering with adaptive $\tau$ and Spectral clustering are shown in Figure \ref{Fig:me}. If referring to Figure \ref{Fig:variable densities}, it seems $K=4$ is a reasonable group number which best reflects the multi-modal property of the marginal density plot. But according to clustering evaluation cretaria including silhouette score and Davies-Bouldin score (Figure \ref{Fig:silouette} and \ref{Fig:db}), both of them showed that $3$ is the optimal cluster number which balances the cluster efficiency and number of clusters from a maximising similarity perspective. Here allow me to presume that a symmetric variance-based clustering algorithm such as $K$-means is not suitable due to the skewness of the whole system. Now we fix $K=3$ and let the algorithm find the optimal location of the cluster center based on the skewness nature of the data, we obtain the $\tau$ parameter in the form of a $(K \times p)$ matrix $[[0.515, 0.448],
[0.222, 0.301],
[0.299, 0.300]]$, corresponding to the blue, green and grey clusters in Figure \ref{Fig:me} respectively. It can also be seen that $K$-means and $K$-expectile algorithms result in different locations of cluster centroids, which lead to different cluster memberships. To better evaluate the performance of $K$-expectile clustering, the result of spectral clustering is using as comparison.

Figure \ref{Fig:clusters} shows the shapes and distribution of the three clusters on the two dimensional space consisting of CRIX and VCRIX. From the plot we can observe that the three clusters of $K$- expectile represents different types of correlation between price and volatility index. The three clusters can be described as 'low-price-low-volatility cluster', 'low-price-high-volatility cluster', and 'positively correlated price and volatility cluster', corresponding to color blue, green and grey. 

it is worth noting that the grey cluster only appears shortly in the end of 2017 and from the end of 2020 till now. Positive correlation between price and volatility of crypto markets means that the volatility and price drives each other in the same direction. Higher price and higher volatility shows an 'exciting' signal other than a 'panic' expectation, this phenomenon mostly occurs in the securities market dominated by individual investors, where increased volatility is a signal of market activation. On the other hand, low-volatility cluster appears in most period of the CC market, which means CC market is highly dominated by instituational investors most of the time. High volatility means unstable market sentiment and high trading volumn, and the green cluster often appears when the price start to change.

\subsection{Application of fixed $\tau$ clustering on image segmentation}
Image segmentation is a technique widely used in image processing, which partition an image into multiple parts sharing similar characteristics. Image segmentation includes separating foreground from background, or clustering regions of pixels based on color or shape. One of the commenly used methods in color-based segmentation is $K$-means clustering. In this case, pixel values are regarded as independent random samples in the 8 bits color space, and divided into $K$ discrete regions which has minimal variances.
The output of $K$- means segmentation can be visualized by converting all the pixels in a group to the color of the cluster centers. 

By applying K-expectile clustering, we expect a more flexible choice of centers and thus a more flexible segmentation output of the image, including an optimising procedure on parameter $\tau$ and two ways to specify $\tau$, which put more weights on group-wised and dimension-wised tail behavior. With $4$ clusters, we set a group-specified $\tau$ as $[0.2, 0.7, 0.1, 0.9]$ to include groups  emphasizes on both left tail and right tail. For $3$ dimensions of RGB valued data, we fixed a parameter $\tau$ as $[0.1,0.8,0.9]$ to involve information on both tails. To test the performance of K-expectile clustering, we take K-means clustering result as benchmark and bring Quantile based clustering results into comparison. 

\begin{figure}[htbp]
		\makebox[\textwidth][c]{\includegraphics[scale=0.5,angle=0]{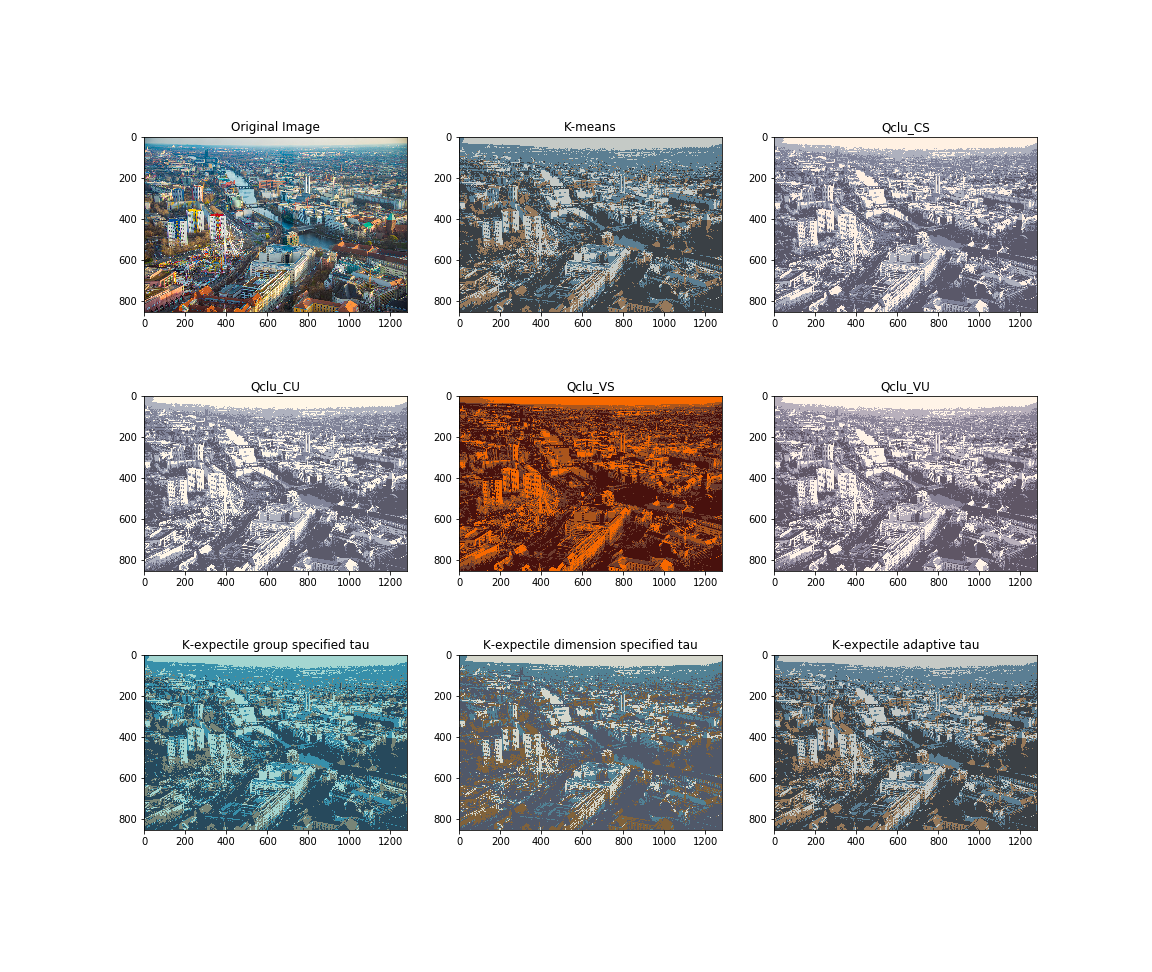}}
		\caption{Image segmentation results using different clustering methods.}
		\label{Fig:Img_overall}
\end{figure}

\begin{figure}[htbp]
		\makebox[\textwidth][c]{\includegraphics[scale=0.5,angle=0]{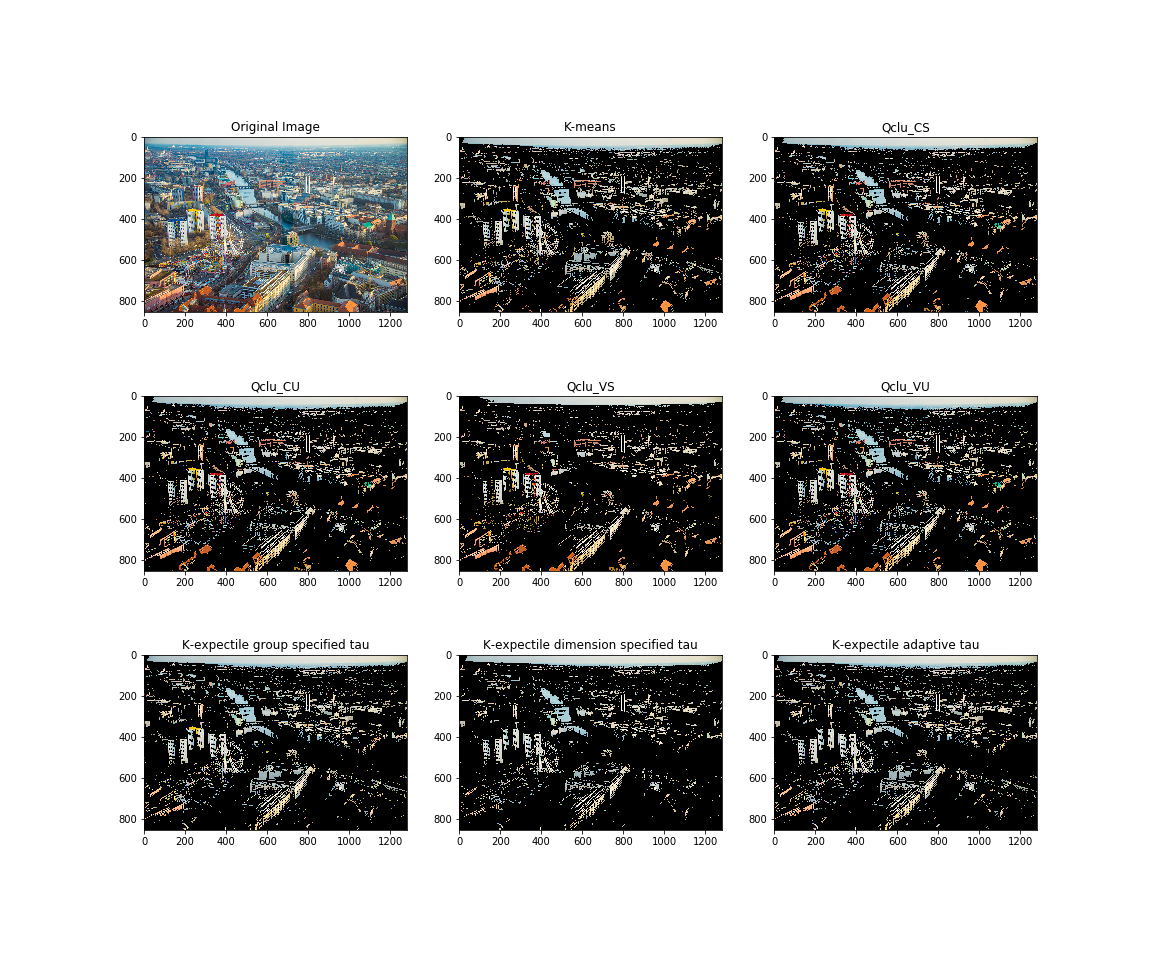}}
		\caption{Segmented image with only one cluster displayed.}
		\label{Fig:Img_cluster}
\end{figure}

The original image is an aerial photo of Berlin, for pre-processing, we transform the image into pixel values in RGB color space, and flatten the data into a two-dimensional array. Figure \ref{Fig:Img_overall} shows the original image and the segmented image, which can be considered as filtered image with 4 color clusters. Important information can be extracted from the image by displaying some clusters and mute others. The subplots showed in Figure \ref{Fig:Img_cluster} are image with only one cluster enabled.

To evaluate the performance of segmentation methods, we use two indices, Mean Square Error (MSE) and Peak to Signal Noise Ration (PSNR). Given an $m\times n$ monochrome image $I$, Mean Square Error measures how much the approximation $K$ differs from it. $MSE$ is defined as: 
\begin{equation*}
MSE = \frac{1}{mn}\sum_{i=1}^{m-1}\sum_{j=1}^{n-1}\{I(i,j)-K(i,j)\}^2.
\end{equation*} 
Peak to Signal Noise Ration is usually used to measure the quality of the compressed image. $PSNR$ is the proportion between maximum attainable powers and the corrupting noise that influence likeness of image. It is defined as following:
\begin{equation*}
PSNR = 10 \operatorname{log}_{10}(\frac{MAX_{I}^{2}}{MSE}),
\end{equation*}
where $MAX_{I}$ is the maximum possible pixel value of the image $I$, which equals to $225$ when the sample is in 8 bits. The higher value of $PSNR$ and the lower value of $MSE$, the better the fitting of the approximated image. 

Table \ref{Tab:image_seg} shows the $MSE$ and $PSNR$ values of segmented image using multiple methods. Although $MSE$ is usually calculated on monochrome data, here we take the average MSE on three RGB dimensions. Moreover, we convert both the original image and the segmented image from RGB data into GRAY and YCrCb color space. From the table, it can be concluded that 1) $K$-expectiles with adaptive $\tau$ perform better compare to both $K$-means and quantile based clustering, 2) with pre-specified parameter $\tau$, in both senario, $K$-expectile clustering gives us a quite moderate result and more flexibility for one to customize the desired segmentation result. 

\begin{table}[htbp]
  \centering
   \begin{tabular}{lrrrrrr}
   	\hline \hline
   	& \multicolumn{2}{c}{GREY} & \multicolumn{2}{c}{YCrCb} & \multicolumn{2}{c}{RGB} \\
   	& \multicolumn{1}{c}{MSE} & \multicolumn{1}{c}{PSNR} & \multicolumn{1}{c}{MSE} & \multicolumn{1}{c}{PSNR} & \multicolumn{1}{c}{MSE} & \multicolumn{1}{c}{PSNR} \\
   	\midrule
   	K-means & 509.18 & 21.06 & 839.12 & 18.89 & 742.53 & 28.96 \\
   	K-expectiles\_vtau & 429.47 & 21.80 & 835.66 & 18.91 & 741.17 & 28.97 \\
   	CS    & 2001.30 & 15.12 & 2886.34 & 13.52 & 2841.61 & 23.14 \\
   	CU    & 2217.68 & 14.67 & 3546.76 & 12.63 & 3639.46 & 22.06 \\
   	VS    & 5338.85 & 10.86 & 2799.57 & 13.66 & 2398.98 & 23.87 \\
   	VU    & 2030.67 & 15.05 & 3332.15 & 12.90 & 3507.04 & 22.22 \\
   	K-expectiles\_gp\_spec\_tau & 519.81 & 20.97 & 1449.73 & 16.51 & 1203.65 & 26.87 \\
   	K-expectiles\_dim\_spec\_tau & 876.31 & 18.70 & 1304.86 & 16.97 & 1214.55 & 26.83 \\
   	\hline \hline
   \end{tabular}%
\caption{Performance of different clustering methods on image segmentation. Data is transformed into RGB, GREY and YCrCb space. Algorithms listed from top to bottom are $K$-means, $K$-expectile with adaptive $\tau$, four modes of Quantile based clustering: Common skewness parameter and Scaled variable-wise, Common skewness parameter and Unscaled variables, Variable-wise skewness parameter and Scaled variable-wise, Variable-wise skewness parameter and Unscaled variables, $K$-expectile with group-specified $\tau$ and $K$-expectile with dimension-specified $\tau$.}
  \label{Tab:image_seg}%
\end{table}%

\begin{figure}[htbp]
  	\makebox[\textwidth][c]{\includegraphics[scale=0.6,angle=0]{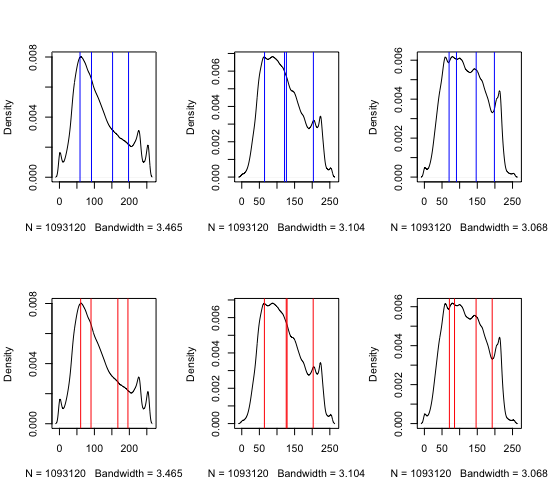}}
  	\caption{Marginal distribution of R-G-B scores and cluster centers. Top: $K$-means. Bottom: $K$-expectile}
  	\label{Fig:Img_marginal}
  \end{figure}

Marginal distribution of R-G-B three dimensional scores has been shown in Figure \ref{Fig:Img_marginal}. Blue and red vertical lines are the locations of $4$ cluster centers of $K$-means and $K$-expectiles with adaptive $\tau$.  The resulting skewness parameter $\tau = [[0.349, 0.469, 0.600],
[0.530,  0.526, 0.675],$\\
$[0.560, 0.450, 0.490],
[0.478, 0.527, 0.475]]$, while for $K$-means, $\tau$ equals to $0.5$ for each cluster and each dimension.  $K$-expectile clustering again shows more adaptibility to adjust to the skewness of the data.

\newpage
\addcontentsline{toc}{section}{References}
\bibliography{literature}


\newpage
\section{APPENDIX A: proofs for Section 2}
In order to proof the convergence of the algorithm, we first recall the objective function:
\begin{align*}
G^{Adaptive}(\tau,\Theta,C,X)&=\sum_{k=1}^{K}\sum_{x_i \in G_k}\sum_{j=1}^{p}d(x_{i j},\tau_{k,j},\theta_{k,j}) \\
&=\sum_{k=1}^{K} \sum_{x_i \in G_k}\sum_{j=1}^{p} \left\lbrace\tau_{k,j}+(1-2\tau_{k,j})  \operatorname{\mathbf{I}}_{\left\lbrace x_{i j}<\theta_{k,j}\right\rbrace } \right\rbrace (x_{i j}-\theta_{k,j})^{2}. \label{eq16}
\end{align*}

Then define
\begin{equation*}
\hat{\theta}_{C(i),j} = \operatorname{arg}\underset{
	\theta_{j}} {\operatorname{min}}\sum_{C(i)=k} \sum_{j=1}^{p} \left\lbrace\tau_{C(i),j}+(1-2\tau_{C(i),j})  \operatorname{\mathbf{I}}_{\left\lbrace x_{ij}<\theta_{j}\right\rbrace } \right\rbrace (x_{ij}-\theta_{j})^{2},
\label{eq17}
\end{equation*}

\begin{equation*}
\hat{\tau}_{C(i),j} = \operatorname{arg}\underset{
	\tau_{j}} {\operatorname{min}}\sum_{C(i)=k} \sum_{j=1}^{p} \left\lbrace\tau_{j}+(1-2\tau_{j})  \operatorname{\mathbf{I}}_{\left\lbrace x_{ij}<\theta_{C(i),j}\right\rbrace } \right\rbrace (x_{ij}-\theta_{C(i),j})^{2} .
\end{equation*}

Let $C_{(i)}^{(t-1)}$ be the previous partition, $\hat{\theta}_{k,j}^{(t-1)} $ and $\hat{\tau}_{k,j}^{(t-1)}$ be previous estimated centroid and $\tau$ parameters, $C_{(i)}^{(t)}$ be the new partition,

\begin{equation*}
G(C_{(i)}^{(t)})\leq \sum_{k=1}^{K} \sum_{C^{(t)} _{(i)}=k}\sum_{j=1}^{p} \left\lbrace\hat{\tau}_{k,j}^{(t-1)} +(1-2\hat{\tau}_{k,j}^{(t-1)} )  \operatorname{\mathbf{I}}_{\left\lbrace x_{ij}<\theta_{j}\right\rbrace } \right\rbrace (x_{ij}-\hat{\theta}_{k,j}^{(t-1)})^{2}.
\end{equation*}

New partition $C_{(i)}^{(t)}$ minimises
$\sum_{k=1}^{K} \sum_{C(i)=k}\sum_{j=1}^{p} \left\lbrace\hat{\tau}_{k,j}^{(t-1)} +(1-2\hat{\tau}_{k,j}^{(t-1)} )  \operatorname{\mathbf{I}}_{\left\lbrace x_{ij}<\theta_{j}\right\rbrace } \right\rbrace (x_{ij}-\hat{\theta}_{k,j}^{(t-1)})^{2} $ over all possible partitions:

\begin{align*}
\sum_{k=1}^{K} \sum_{C^{(t)} _{(i)}=k}\sum_{j=1}^{p} \left\lbrace\hat{\tau}_{k,j}^{(t-1)} +(1-2\hat{\tau}_{k,j}^{(t-1)} )  \operatorname{\mathbf{I}}_{\left\lbrace x_{ij}<\theta_{j}\right\rbrace } \right\rbrace (x_{ij}-\hat{\theta}_{k,j}^{(t-1)})^{2} \\
 \leq   \underbrace{\sum_{k=1}^{K} \sum_{C^{(t-1)} _{(i)}=k}\sum_{j=1}^{p} \left\lbrace\hat{\tau}_{k,j}^{(t-1)} +(1-2\hat{\tau}_{k,j}^{(t-1)} )  \operatorname{\mathbf{I}}_{\left\lbrace x_{ij}<\theta_{j}\right\rbrace } \right\rbrace (x_{ij}-\hat{\theta}_{k,j}^{(t-1)})^{2}}_{G(C_{(i)}^{(t-1)})}.
\end{align*}

Hence,$G(C_{(i)}^{(t)})\leq G(C_{(i)}^{(t-1)})$.

\newpage

\section{Figures}

\begin{figure}[htbp]
	\begin{center}
		\includegraphics[scale=0.4,angle=0]{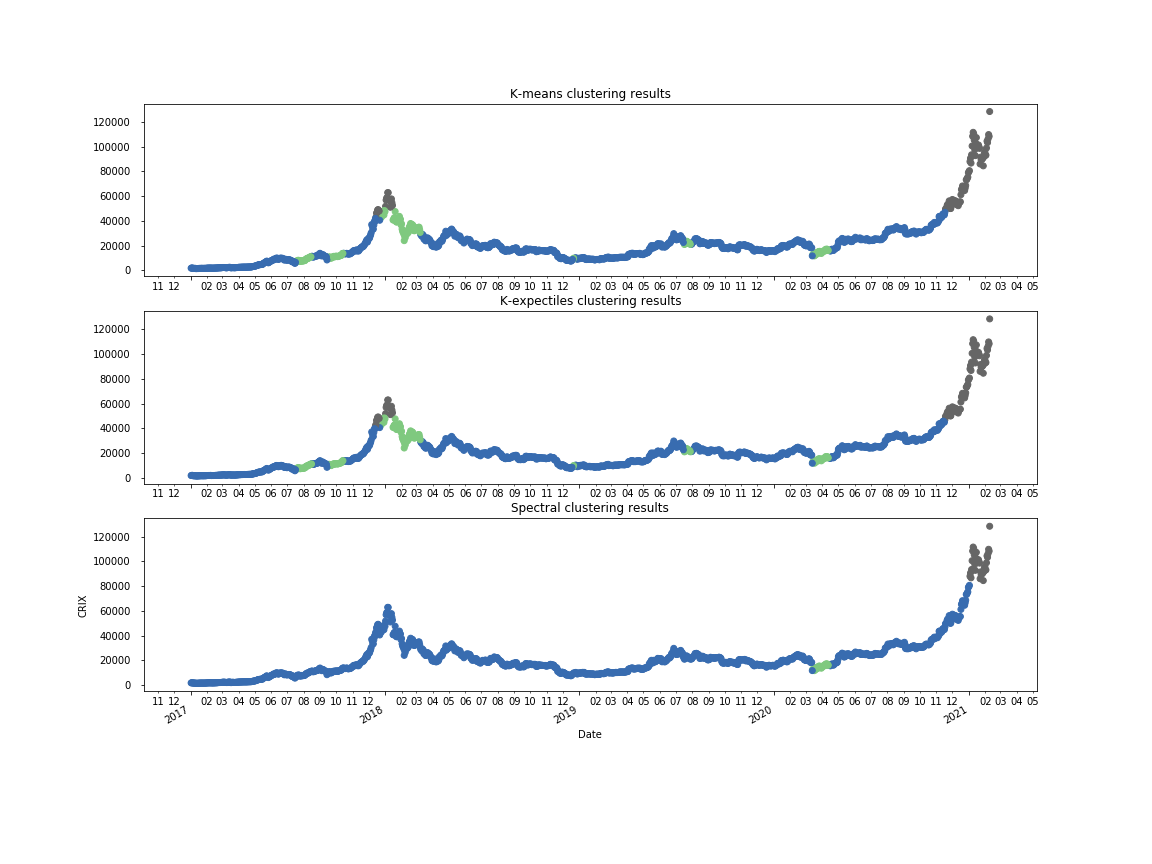}
		\caption{Clustering results of $K$-means, $K$-expectiles and Spectral clustering on CRIX .Clusters are shown in different colours.}\includegraphics[scale=0.05]{qletlogo_tr.png}\href{https://github.com/QuantLet/KEC/tree/master/KEC_applications}{KEC\_application}
		\label{Fig:CRIX}
	\end{center}
\end{figure}

\begin{figure}[htbp]
	\begin{center}
		\includegraphics[scale=0.4,angle=0]{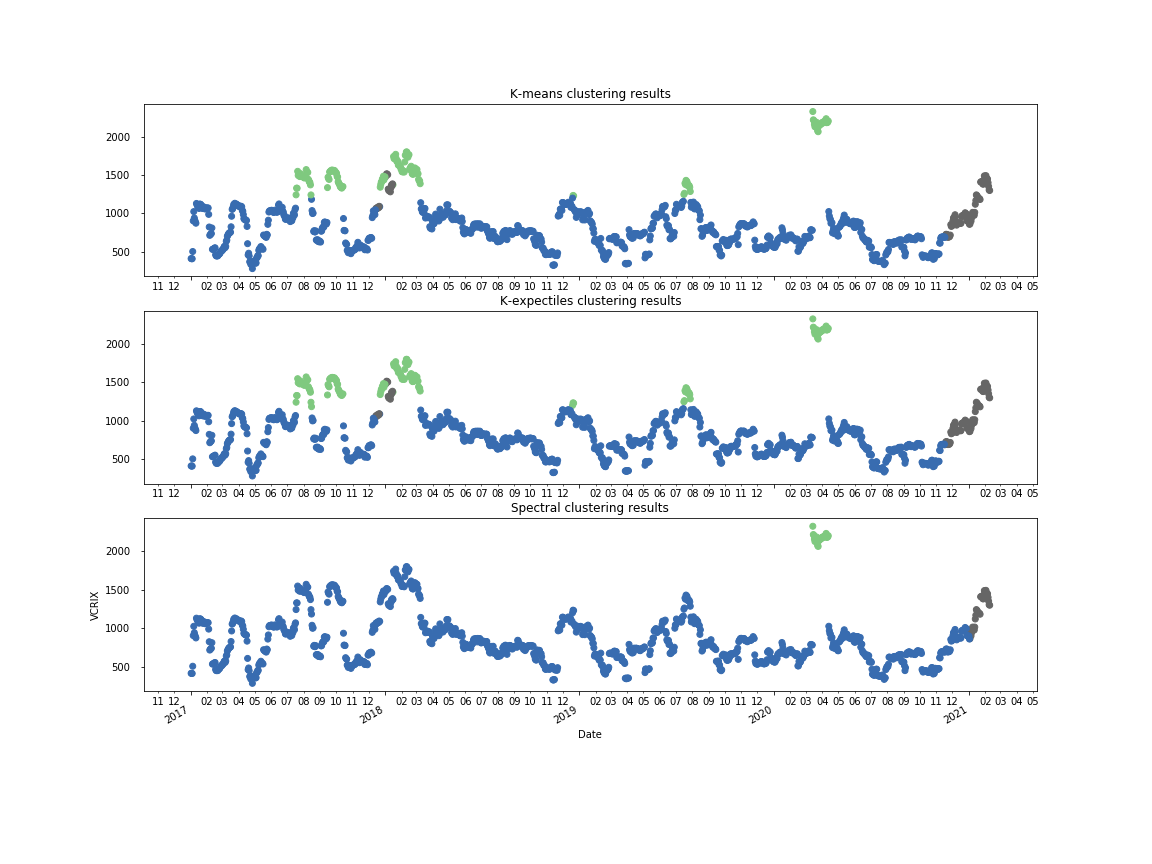}
		\caption{Clustering results of $K$-means, $K$-expectiles and Spectral clustering on VCRIX .Clusters are shown in different colours.}\includegraphics[scale=0.05]{qletlogo_tr.png}\href{https://github.com/QuantLet/KEC/tree/master/KEC_applications}{KEC\_application}
		\label{Fig:VCRIX}
	\end{center}
\end{figure}

\newpage
\section{Tables}

 \begin{table}[htbp]

 	\centering

 	\caption{Sample1: Simulation results of Gaussian clusters}\includegraphics[scale=0.05]{qletlogo_tr.png}\href{https://github.com/QuantLet/KEC/tree/master/KEC_simulations}{KEC\_simulations}

 	\begin{tabular}{lcccccc}

 		\hline\hline

 		\multicolumn{4}{c}{$n$ =1500}   & \multicolumn{3}{c}{$n$ = 300} \\

 		\midrule

 		& \multicolumn{1}{l}{$p$=10} & \multicolumn{1}{l}{$p$=50} & \multicolumn{1}{l}{$p$=100} & \multicolumn{1}{l}{$p$=10} & \multicolumn{1}{l}{$p$=50} & \multicolumn{1}{l}{$p$=100} \\

 		& \multicolumn{1}{c}{ARI} & \multicolumn{1}{c}{ARI} & \multicolumn{1}{c}{ARI} & \multicolumn{1}{c}{ARI} & \multicolumn{1}{c}{ARI} & \multicolumn{1}{c}{ARI} \\

 		\midrule

 		K-expectiles\_vtau & 99.36 & 99.60 & 99.87 & 97.00 & 97.99 & 97.99 \\

 		K-means & 99.36 & 99.60 & 99.60 & 97.00 & 97.99 & 97.99 \\

 		Spectral &       & 31.22 & 86.74 &       & 27.48 & 85.03 \\

 		h-ward & 99.20 & 99.60 & 99.87 & 93.54 & 97.99 & 97.99 \\

 		CS    & 99.24 & 99.60 & 99.87 & 96.61 & 97.99 & 97.99 \\

 		CU    & 99.24 & 99.60 & 99.87 & 96.61 & 97.99 & 97.99 \\

 		VS    & 99.28 & 99.60 & 99.87 & 96.03 & 97.99 & 97.99 \\

 		VU    & 99.20 & 99.60 & 99.87 & 96.61 & 97.99 & 97.99 \\

 		\hline\hline

 	\end{tabular}%

 	\label{tab:sim1}%

 \end{table}%


\begin{table}[htbp]

	\centering

	\caption{Sample2: Simulation results of Asymmetric normal clusters}\includegraphics[scale=0.05]{qletlogo_tr.png}\href{https://github.com/QuantLet/KEC/tree/master/KEC_simulations}{KEC\_simulations}

	\begin{tabular}{lcccccc}

		\hline\hline
		\multicolumn{4}{c}{$n$ =1500}   & \multicolumn{3}{c}{$n$ = 300} \\
		
		\midrule
		
		& \multicolumn{1}{l}{$p$=10} & \multicolumn{1}{l}{$p$=50} & \multicolumn{1}{l}{$p$=100} & \multicolumn{1}{l}{$p$=10} & \multicolumn{1}{l}{$p$=50} & \multicolumn{1}{l}{$p$=100} \\
		& \multicolumn{1}{c}{ARI} & \multicolumn{1}{c}{ARI} & \multicolumn{1}{c}{ARI} & \multicolumn{1}{c}{ARI} & \multicolumn{1}{c}{ARI} & \multicolumn{1}{c}{ARI} \\

		\midrule

		K-expectiles\_vtau & 93.22 & 99.60 & 99.60 & 92.20 & 97.99 & 97.99 \\

		K-means & 91.19 & 99.59 & 99.60 & 81.70 & 97.99 & 97.99 \\

		Spectral &       & -0.02 &       &       &       &  \\

		h-ward & 77.19 & 99.52 & 99.60 & 76.98 & 97.01 & 97.99 \\

		CS    & 86.61 & 99.58 & 99.60 & 88.98 & 97.99 & 71.74 \\

		CU    & 80.73 & 99.28 & 94.70 & 72.76 & 93.36 & 76.27 \\

		VS    & 88.86 & 99.59 & 99.57 & 93.16 & 91.71 & 97.99 \\

		VU    & 85.74 & 99.55 & 99.60 & 80.41 & 93.73 & 97.99 \\

		\hline\hline

	\end{tabular}%

	\label{tab:sim2}%

\end{table}%


\begin{table}[htbp]

	\centering

	\caption{Sample 3: Simulation results of $Beta$-distributed clusters}\includegraphics[scale=0.05]{qletlogo_tr.png}\href{https://github.com/QuantLet/KEC/tree/master/KEC_simulations}{KEC\_simulations}

	\begin{tabular}{lcccccc}

		\hline\hline
		\multicolumn{4}{c}{$n$ =1500}   & \multicolumn{3}{c}{$n$ = 300} \\
		
		\midrule
		
		& \multicolumn{1}{l}{$p$=10} & \multicolumn{1}{l}{$p$=50} & \multicolumn{1}{l}{$p$=100} & \multicolumn{1}{l}{$p$=10} & \multicolumn{1}{l}{$p$=50} & \multicolumn{1}{l}{$p$=100} \\

		& \multicolumn{1}{c}{ARI} & \multicolumn{1}{c}{ARI} & \multicolumn{1}{c}{ARI} & \multicolumn{1}{c}{ARI} & \multicolumn{1}{c}{ARI} & \multicolumn{1}{c}{ARI} \\

		\midrule

		K-expectiles\_vtau & 94.04 & 99.60 & 99.60  & 93.17 & 97.99 & 97.99 \\

		K-means & 94.79 & 99.60 & 99.60 & 93.16 & 97.99 & 97.99 \\

		Spectral & 93.63 & 99.60 & 99.60 & 93.17 &       &        \\

		h-ward & 94.80 & 99.60 & 99.60 & 88.88 & 97.99 & 97.99 \\

		CS    & 68.92 & 96.89 & 82.52 & 92.20 & 97.99 & 65.54 \\

		CU    & 68.14 & 94.08 & 73.26 & 92.17 & 93.36 & 65.62 \\

		VS    & 93.28 & 97.47 & 79.84 & 91.98 & 91.71 & 62.46 \\

		VU    & 94.03 & 94.69 & 73.28 & 92.56 & 93.73 & 46.90 \\

		\hline\hline
	\end{tabular}%

	\label{tab:sim3}%

\end{table}%


\begin{table}[htbp]

	\centering

	\caption{Sample 4-1: Simulation results of generalized $t$-distributed clusters}\includegraphics[scale=0.05]{qletlogo_tr.png}\href{https://github.com/QuantLet/KEC/tree/master/KEC_simulations}{KEC\_simulations}
	
	\begin{tabular}{lcccccc}
		\hline\hline
		& \multicolumn{3}{c}{n =1500} & \multicolumn{3}{c}{n = 300} \\
		\midrule
		& \multicolumn{1}{c}{p=2} & \multicolumn{1}{c}{p=10} & \multicolumn{1}{c}{p=50} & \multicolumn{1}{c}{p=2} & \multicolumn{1}{c}{p=10} & \multicolumn{1}{c}{p=50} \\
		& \multicolumn{1}{c}{ARI} & \multicolumn{1}{c}{ARI} & \multicolumn{1}{c}{ARI} & \multicolumn{1}{c}{ARI} & \multicolumn{1}{c}{ARI} & \multicolumn{1}{c}{ARI} \\
		\midrule
		K-expectiles\_vtau & 96.50 & 97.99 & 97.99 & 95.10 & 97.99 & 97.99 \\
		K-means & 96.26 & 97.99 & 97.99 & 94.80 & 97.99 & 97.99 \\
		Spectral & 96.21 & 26.31 &       & 94.81 & 93.10 & 92.43 \\
		h-ward & 96.24 & 97.99 & 97.99 & 94.29 & 97.99 & 97.99 \\
		CS    & 96.48 & 97.99 & 97.99 & 95.01 & 97.99 & 97.99 \\
		CU    & 96.08 & 97.99 & 97.99 & 94.57 & 97.99 & 97.99 \\
		VS    & 96.48 & 97.99 & 97.99 & 95.10 & 97.99 & 97.99 \\
		VU    & 96.07 & 97.99 & 97.99 & 94.57 & 97.99 & 97.99 \\
		\hline\hline
	\end{tabular}%

	\label{tab:sim4-1}%

\end{table}%


\begin{table}[htbp]

	\centering

	\caption{Sample 4-2: Simulation results of $F$-distributed clusters}\includegraphics[scale=0.05]{qletlogo_tr.png}\href{https://github.com/QuantLet/KEC/tree/master/KEC_simulations}{KEC\_simulations}

    \begin{tabular}{lcccccc}
    	\hline\hline
          & \multicolumn{3}{c}{n =1500} & \multicolumn{3}{c}{n = 300} \\
    \midrule
          & \multicolumn{1}{c}{p=2} & \multicolumn{1}{c}{p=10} & \multicolumn{1}{c}{p=50} & \multicolumn{1}{c}{p=2} & \multicolumn{1}{c}{p=10} & \multicolumn{1}{c}{p=50} \\
          & \multicolumn{1}{c}{ARI} & \multicolumn{1}{c}{ARI} & \multicolumn{1}{c}{ARI} & \multicolumn{1}{c}{ARI} & \multicolumn{1}{c}{ARI} & \multicolumn{1}{c}{ARI} \\
    \midrule
    K-expectiles\_vtau & 95.80 & 99.60 & 99.60 & 94.58 & 99.60 & 99.60 \\
    K-means & 95.19 & 99.60 & 99.60 & 94.01 & 99.60 & 99.60 \\
    Spectral & 94.89 & 26.31 &       & 93.82 &       &  \\
    h-ward & 96.82 & 99.60 & 99.60 & 95.25 & 99.60& 99.60 \\
    CS    & 97.96 & 99.60 & 99.60 & 96.03 & 99.60 & 99.60 \\
    CU    & 95.42 & 99.60 & 99.60 & 94.19 & 99.60 & 99.60 \\
    VS    & 97.72 & 99.60 & 99.60 & 95.64 & 99.60 & 99.60 \\
    VU    & 95.44 & 99.60 & 99.60 & 94.19 & 99.60 & 99.60 \\
    \hline\hline
    \end{tabular}%
	\label{tab:sim4-2}%

\end{table}%


\newpage

\end{document}